# Femtosecond laser pulse induced Coulomb explosion

[1]R.ANNOU and V.K.TRIPATHI

Physics Department, Indian Institute of Technology, New-Delhi 16, INDIA
[1]*Permanent address*: Faculty of physics, University of Sciences and Technology Houari Boumediene, Algiers, ALGERIA.

## Abstract

Coulomb explosion is revisited on the basis of explosion theory. The characteristics of the Coulomb explosion such as the ion kinetic energy, the energy distribution function, the laser absorption length along with the neutron yield are determined. The model explains reasonably well the experimental results.

High intensity femto-second laser pulse matter interaction has revealed a new host of phenomena with tremendous potential applications. During laser-molecule and laser-cluster interaction, electrons are stripped off the particulates that explode subsequently under the influence of the stored electrostatic energy. The laser-cluster interaction exhibits characteristic features, which allows one to rank it as a transition between laser-gas and laser-solid interactions. The Coulomb explosion generates bright keV x-ray photons, highly energetic electrons and very fast ions that are emitted along the laser polarization for exploding molecules, but are isotropically emitted for clusters[1-8]. Many applications based on Coulomb explosion have been pointed out, such as i/ Coulomb explosion imaging, where a fast triatomic molecular ion impinges on a thin foil that strip off the electrons, leading to the repulsion of the positively charged remnants and freezing of the velocities up to an imaging detector where the initial molecular structure may be traced back, ii/ source of soft and hard x-rays, the x-ray yield may be cluster size and density of clusters dependent, iii/ source of pulsed and fast neutrons generated in appropriate gas jets, like those containing deuterium (**D**) or tritium (**T**). The neutrons are the result of D-D or D-T interactions of fast elements ejected from different clusters. The neutron yield depends on cluster size and laser parameters, iv/ or simply a way for achieving fusion, since exploding clusters provide sufficient kinetic energy for deuterons to trigger a fusion reaction. An alternative would be the use of the energetic ions for heating a pre-compressed D-T pellet up to ignition[9-12].

Depending on the cluster size $r_c$ and the electron excursion $\xi_e$ under the influence of the laser electric field, the expansion of the ionized cluster may be seen as a Coulomb explosion ($r_c \ll \xi_e$) or an ambipolar expansion ($r_c \gg \xi_e$). The theories developed so far to explain all the features of the Coulomb explosion are based on the second category. In this letter however, we believe since we are coping with a Coulomb explosion, it is appropriate to tackle the problem through the explosion theory rather than the sudden expansion of a cloud. As a matter of fact, we retrieve almost all the experimental facts pointed out in Coulomb explosion **experiments.**

Consider a gas of numerical density $n_a$, comprising $n_c$ clusters per unit volume. The cluster has numerical density $n_0$ and radius $r_c$. A laser pulse interacts with the gas,

$$\vec{E}_L = A(t,z)\exp(-i(\omega t - kz))\hat{x}. \tag{1}$$

The laser field tunnel ionizes the atoms of the cluster and imparts quiver velocity to the freed electrons,

$$\vec{v}_L = \frac{e\vec{E}_L}{mi\omega}, \tag{2}$$

with excursion being $x_e = \frac{eA}{m\omega^2}$. We assume a high laser intensity to allow for $x_e \gg r_c$, consequently electrons move out of the cluster forming an electron cloud. During this time, the positive ions are immobile over the entire sphere of radius $r_c$. Let's recall, that to bring about a number of positive charges from infinity to form one system, we should perform a work. If the cluster is not maintained by a balancing force, it will restitute the stored or the electrostatic energy back to the surrounding environment. To calculate the stored energy in the cluster of positive ions aggregating in a sphere, we consider the work done in bringing the positive charges from infinity to form a crystal, that is given by,

$$W = (3 + \frac{6}{\sqrt{2}} + \frac{4}{\sqrt{3}})\frac{2e^2}{a}\sum_{n=1}^{N}\frac{1}{n}, \tag{3}$$

where a is the inter-ion distance. Unfortunately, this form is not tractable since the harmonic infinite series is divergent. We approximate then, the energy stored by this finite number of ions, by the energy stored in a positively charged sphere of radius $r_c$, namely,

$$W = \frac{3}{5}\frac{Q^2}{r_c} = \frac{16\pi^2}{15}n_0^2 e^2 Z^2 r_c^5. \tag{4}$$

The energy density being given by,

$$w = \frac{4\pi}{5}n_0^2 e^2 Z^2 r_c^2. \tag{5}$$

The so-called Coulomb explosion pressure is proportional to the second power of the cluster size,

$$P_{coul} = \frac{\left|-\partial W/\partial r_c\right|}{4\pi r_c^2} = \frac{4\pi}{3}n_0^2 e^2 Z^2 r_c^2. \tag{6}$$

**On the other the hydrodynamic pressure** $P_e = n_e T_e$ **scales as** $1/r_c^3$; **consequently, since** $P_{coul}/P_e \propto r_c^5$, for big clusters the explosion is driven by the electrostatic energy stored in the sphere rather than by the space-charge field. When all electrons have been removed, the energy stored in the positive ions sphere, has to be restituted in an explosively manner. The outer surface ions will move radially in the electron cloud with a velocity $\vec{u}$, generating subsequently a shock wave in that cloud. The velocity of the contact surface (CS) $\vec{u}$ will be approximated by the shock front (SF) velocity $\vec{D}$, viz., $\vec{u} \approx \vec{D}$. The shock front is a transition between two states of the electron gas (indexed by 1 and 0), where the following relations valid for strong shocks hold (cf. Ref.(13)),

$$n_1 \approx n_0 \frac{\gamma+1}{\gamma-1};$$

$$p_1 \approx \frac{2}{\gamma+1} n_0 D^2; \quad (7)$$

$$v_1 \approx \frac{2}{\gamma+1} D;$$

the electron gas may be considered ideal with the non-ideal character brought through an appropriate choice of the adiabatic exponent γ. One may consider also, in view of improving the model, the degenerate plasma equation of state. By virtue of the similarity method of gas-dynamics, we know that the only non-dimensional variable would be,

$$x = r \left( \frac{r_0}{W t^2} \right)^{1/5}. \quad (8)$$

Then the shock front is propagating, and its radius scales as,

$$R(t) = x_0 \left( \frac{W}{r_0} \right)^{1/5} t^{2/5}, \quad (9)$$

where $\xi_0$ is a constant that is given by (c.f.Ref.(13)), $x_0 = \left( \frac{75}{16} \frac{(\gamma-1)(\gamma+1)^2}{(3\gamma-1)} \right)^{1/5}$.

The velocity of the (SF) scales at R>>$r_c$, as

$$D(R) = \frac{2}{5}x_0^{5/2}\left(\frac{W}{r_0}\right)^{1/2} R^{-3/2}. \tag{10}$$

The velocity of the SF decreases with distance. Moreover, may be pointed out that Eq.(10) is valid in particular for $R=r_c$, thus,

$$D(r_c) = \frac{2}{5}x_0^{5/2}\left(\frac{W}{r_0}\right)^{1/2} r_c^{-3/2} = \frac{8\pi}{5\sqrt{15}}\left(\frac{x_0^5}{r_0}\right)^{1/2} n_0 eZ r_c. \tag{11}$$

Equation (11), reveals that ions on the outer surface of the cluster get higher kinetic energy. For the parameters of Ditmire *et al's* experiment, $n_0 = 3\times 10^{22}$ cm$^{-3}$, $r_c = 50$Å, $\varepsilon_m = mD^2(r_c)/2 = 5.4$ keV; 2.16 MeV for Z=1; 20.

Let us then derive the energy distribution of the ejected ions. We consider a thin shell of radius $r_c$. The number of ions contained in this shell having energies in the range $\varepsilon$ and $\varepsilon + d\varepsilon$, is given by

$$dN = 4\pi n_0 r_c^2 dr_c = f(\varepsilon)d\varepsilon. \tag{12}$$

Consequently the distribution function is as follows,

$$f(\varepsilon) = 4\pi n_0 r_c^2 \frac{dr_c}{d\varepsilon}. \tag{13}$$

However, from Eq.(11) the kinetic energy is derived, i.e., $\varepsilon = \frac{1}{2}mD^2$, to obtain

$$\frac{d\varepsilon}{dr_c} = m\left(\frac{8\pi}{5\sqrt{15}}\right)^2 \frac{x_0^5}{r_0}(n_0 eZ)^2 \, r_c. \tag{14}$$

Finally, the distribution function reads as,

$$\frac{f}{f^*} = \sqrt{\varepsilon}, \tag{15}$$

where $f^{*1/3} = \dfrac{5\sqrt{15}}{2^{13/6}\pi^{2/3}e\varepsilon_0^{15/6}Zn_0^{1/6}}$.

This energy distribution function has already been assumed in Ref.(12). One should keep in mind that beyond the maximum energy $\varepsilon_m$, $f(\varepsilon)$ should vanish. The maximum energy available for ions being $\varepsilon_m = mD^2(r_c)/2 = AZ^2 r_c^2$, with $A = 32\pi^2 n_0 e^2 \varepsilon_0^5 / 375$. It is revealed that the maximum kinetic energy is a quadratic function of the cluster size and the ion charge, in accordance with earlier experimental work[3,7].

The average kinetic energy is given straightforwardly by,

$$<\varepsilon> = \dfrac{\int_0^{\varepsilon_m} \varepsilon f(\varepsilon)d\varepsilon}{\int_0^{\varepsilon_m} f(\varepsilon)d\varepsilon} = \dfrac{3}{5}\varepsilon_m. \tag{16}$$

For a power-law cluster size-distribution, namely, $g(r_c) = C_p r_c^{-p}$, one may obtain a modified energy distribution such as (c.f. Ref.(12)),

$$f(\varepsilon) \propto C_p \dfrac{r_{c2}^{1-p} - r_{c1}^{1-p}}{1-p}\sqrt{\varepsilon}, \tag{17}$$

where $r_{c1}$ and $r_{c2}$ are the minimum and maximum radii of the clusters.

By stripping off the electrons, the laser pulse acted as an operator that rubes matches against a rugous surface to ignite a combustion reaction, which in turn provides more energy. To calculate the absorption length, one should evaluate the energy given by the laser to the clouds of electrons in all the clusters.

By virtue of Eq.(2), we calculate the energy provided by the laser pulse to each electron,

$$U_P = \dfrac{1}{2}m_e v_L^2 = \dfrac{e^2 E_L^2}{2 m_e \omega^2}. \tag{18}$$

For a cluster, the energy turns out to be,

$$U_{PC} = \dfrac{4\pi}{3}\dfrac{n_0 e^2 E_L^2}{2 m_e \omega^2} r_c^3. \tag{19}$$

However, in a cylinder of a length $L_{//}$ corresponding to the absorption length, and a base of surface area S, the pulse encounters $N_c = L_{//} S n_c$ clusters hence the energy delivered by the laser to the total number of electrons is finally given by,

$$U_{PCT} = L_{//} S n_c U_{PC.}$$ (20)

We keep in mind that the energy of the pulse is $I_L S \tau_L$, and then one may determine the absorption length if no dissipation effect is taken into account by equating the energies to get,

$$L_{//} = \frac{I_L \tau_L}{n_c U_{PC.}}.$$ (21)

This length is independent of the laser intensity because it is calculated for a maximum electron depletion of the cluster. For the parameters, $n_0 = 10^{22}$ cm$^{-3}$, $r_c$=50Å, $I_L=10^{16}$W/cm$^2$, $\tau_L$ =35fs, if one takes $n_c=10^{17}$ cm$^{-3}$, the attenuation length turns out to be $L_{//} = 33$ μm. This length is shorter than the length of the gas column containing clusters. Hence experimental value of 90% laser energy absorption is consistent with our model. Full absorption of laser energy does not occur as a reduced intensity laser pulse can not cause tunnel ionization of atoms and propagates unaffected after its level has fallen below the threshold for tunnel ionization.

Coulomb explosion generates a hot ions component in addition to the cold one already present in the plasma. In the case of deuterium plasma jets, the hot deuterons would collide with others of the surrounding media, to achieve fusion reactions, which give rise consequently to neutrons. To calculate the number of neutrons generated by the laser pulse, we first determine the effective number of energetic ions undergoing a thermonuclear reaction. The number of energetic ions produced via cluster Coulomb explosion by a laser pulse of spot size $r_0$ is given by,

$$N_{ion} = \frac{4\pi}{3} n_0 n_c \, r_c^3 \, \pi \, r_0^2 \, L_{//}$$ (22)

the fraction of those ions that undergo fusion in the surrounding gas of the size L is $L n_a \sigma_f$ for each reaction, where $\sigma_f$ is the fusion cross-section of deuterium. The total number of fusion reactions in the plasma is,

$$N_f = N_{ion} L_\perp n_a (\sigma_f + \sigma_{f'})$$ (23)

and the fusion energy released is given by,

$$W_f = N_{ion} L_\perp n_a (\sigma_f e_f + \sigma_{f'} e_{f'}) \tag{24}$$

where $e_f$ and $e_{f'}$ are the fusion energy released per each fusion reaction. Total energy of the laser pulse is $W_L = I_L t_L \pi r_0^2$, and then the neutron yield is,

$$n_n = \frac{N_{ion} L_\perp n_a \sigma_f}{W_L} \text{ Neutron/laser energy} \tag{25}$$

For $r_c$=50Å, $n_0$=$10^{22}$cm$^{-3}$, $\sigma$=5x$10^{-28}$cm$^{-2}$ (at 2.5 kev deuterium energy), $n_a$=$10^{19}$ cm$^{-3}$, $L$ =100μm, $n_n$=0.44x$10^5$ neutron per joule of laser energy.

To conclude, during the interaction of high intensity femto-second laser pulses with clusters, the electrons execute large excursions, leaving the positive cluster charge uncompensated. Under the effect of the stored electrostatic energy, large clusters explode releasing energetic ions. The calculations reveal[14] that ions on the outer surface of the cluster get higher kinetic energy, and that the energy is distributed according to a square rooted law. Coulomb explosion is triggered by the laser pulse; hence the laser energy is absorbed on a length that has been determined in favor of experimental facts. In case deuterium gas jets are considered, the release of energetic ions may trigger fusion reactions with the surrounding low energy ions, with a subsequent neutron production. Proton release should be also expected.

**Acknowledgment**

This work is supported by USTHB (Algeria,2003) and in part by IITDelhi (India).